\documentclass[sigconf]{acmart}

\usepackage{graphicx} 
\usepackage{natbib}  
\usepackage{caption} 
\usepackage[utf8]{inputenc} 
\usepackage[T1]{fontenc}    
\usepackage{url}            
\usepackage{float}
\usepackage{adjustbox}
\usepackage{booktabs} 
\usepackage{graphicx}
\usepackage{multirow}
\usepackage{ragged2e}
\usepackage{amsmath}
\usepackage[ruled,vlined,linesnumbered]{algorithm2e}
\usepackage{amsfonts}       
\usepackage{nicefrac}       
\usepackage{microtype}      
\usepackage{enumitem}
\usepackage{xcolor,colortbl}
\usepackage[ruled,vlined,linesnumbered]{algorithm2e}
\usepackage{amsfonts}       
\usepackage{nicefrac}       
\usepackage{microtype}      
\usepackage{enumitem}

\definecolor{lightgray}{rgb}{0.83, 0.83, 0.83}
\definecolor{darkgray}{rgb}{0.5, 0.5, 0.5}
\definecolor{mediumgray}{rgb}{0.66, 0.66, 0.66}

\AtBeginDocument{%
  \providecommand\BibTeX{{%
    \normalfont B\kern-0.5em{\scshape i\kern-0.25em b}\kern-0.8em\TeX}}}

\copyrightyear{2021} 
\acmYear{2021} 
\setcopyright{rightsretained} 
\setcopyright{none} 
\acmConference[AIES '21]{Proceedings of the 2021 AAAI/ACM Conference on AI, Ethics, and Society}{May 19--21, 2021}{Virtual Event, USA}
\acmBooktitle{Proceedings of the 2021 AAAI/ACM Conference on AI, Ethics, and Society (AIES '21), May 19--21, 2021, Virtual Event, USA}\acmDOI{10.1145/3461702.3462561}
\acmISBN{978-1-4503-8473-5/21/05}

\begin{document}

\title{Disparate Impact of Artificial Intelligence Bias in Ridehailing Economy's Price Discrimination Algorithms}

\author{Akshat Pandey}
\affiliation{%
  \institution{George Washington University}
  \institution{Department of Computer Science}
}
\email{apandey0123@gwu.edu}

\author{Aylin Caliskan}
\affiliation{%
  \institution{George Washington University}
  \institution{Department of Computer Science}
  \institution{Institute for Data, Democracy \& Politics}
  }
\email{aylin@gwu.edu}

\renewcommand{\shortauthors}{Pandey and Caliskan}

\begin{abstract}
Ridehailing applications that collect mobility data from individuals to inform smart city planning predict each trip's fare pricing with automated algorithms that rely on artificial intelligence (AI). This type of AI algorithm, namely a price discrimination algorithm, is widely used in the industry's black box systems for dynamic individualized pricing. Lacking transparency, studying such AI systems for fairness and disparate impact has not been possible without access to data used in generating the outcomes of price discrimination algorithms. Recently, in an effort to enhance transparency in city planning, the city of Chicago regulation mandated that transportation providers publish anonymized data on ridehailing. As a result, we present the first large-scale measurement of the disparate impact of price discrimination algorithms used by ridehailing applications.  

The application of random effects models from the meta-analysis literature combines the city-level effects of AI bias on fare pricing from census tract attributes, aggregated from the American Community Survey. An analysis of 100 million ridehailing samples from the city of Chicago indicates a significant disparate impact in fare pricing of neighborhoods due to AI bias learned from ridehailing utilization patterns associated with demographic attributes. Neighborhoods with larger non-white populations, higher poverty levels, younger residents, and high education levels are significantly associated with higher fare prices, with combined effect sizes, measured in Cohen's $d$, of $-0.32$, $-0.28$, $0.69$, and $0.24$  for each demographic, respectively. Further, our methods hold promise for identifying and addressing the sources of disparate impact in AI algorithms learning from datasets that contain U.S. geolocations. 

\end{abstract}



\maketitle
\fancyhead{}

\section{Introduction}

Ridehailing applications are artificial intelligence (AI) based services that allow users to order rides to a specified location from the convenience of their phone.
Transportation network companies, such as Uber and Lyft, provide these services that can be found all over the world.
In major U.S. cities, $21\%$ of adults have used ridehailing services, and in the first five years of adoption, beginning in $2009$, there were a total of $250$ million estimated global users~\cite{clewlow1}.
Unlike traditional taxi services, fare prices for ridehailing services are dynamic, calculated automatically using both the length of the requested trip as well as the demand and relative supply for ridehailing services in the area~\cite{uber3, uber1, chen1}.
Potential AI bias in ridehailing applications, embedded in the algorithm and data, is a concern because of its possible disparate effect on people as well as the implications it has on the future of smart city automation and resource allocation.

In $2017$, Uber launched a service called ``Uber Movement'', a platform on which cities can purchase and access transportation data collected by Uber with the goal of meeting future ``smart city'' aspirations~\cite{attoh1}. 
The lack of extensive algorithmic transparency and the black box nature of ridehailing fare pricing algorithms leads to apprehension of whether they may be unintentionally exhibiting AI bias towards riders based on demographics or their neighborhoods.
If there is bias in the algorithms and data being used in the development of smart cities, it may propagate outside of ridehailing platforms through data-driven policy making and automated resource allocation.
Questions have been raised towards ridehailing providers' driverless car aspirations as well~\cite{benanav1}.
In addition to the economic strain placed on their workers~\cite{attoh1, benanav1}, automating smart city data collection by removing human drivers from the gig economy is concerning given the black box nature of dynamic pricing and supply allocation algorithms. 
As a result, it is still unclear how ridehailing providers impact traffic, supply and demand, and what data is being used to make automated pricing and supply allocation decisions.

In the context of price discrimination in ridehailing, we define AI bias as the difference in algorithmic outcomes for fare pricing associated with demographic characteristics of riders' neighborhoods.
Such unintentional difference in outcomes, or disparate impact, is the current standard for legal determination of discrimination in the United States~\cite{feldman1}.
It was introduced in the case of Griggs vs. Duke Power Co.~\cite{griggs1}, which found African American employees were disparately impacted by employment exams, though unintentionally.
AI systems have been known to adopt policies that display demographic disparity as well \cite{lee2018detecting}. 
The disparity has been documented in algorithmic solutions to job candidate selection, online advertisements, recidivism prediction, and location-based pricing for college test preparation services~\cite{barocas1, datta2015automated, turner_lee1, sweeney2013discrimination, mehrabi1, angwin1}.
AI has also been shown to learn implicit biases documented by social psychologists, such as racial, gender, and other cultural biases, through the statistical regularities of language \cite{caliskan2017semantics}, resulting in biased data representations that may propagate to downstream applications such as automated essay grading, university admissions, or resume screening for job candidate assessment \cite{raghavan2020mitigating}. 

Vendors of job candidate selection software have avoided claims of intentional discrimination by ensuring that their models do not contain legally protected attributes, such as race, religion, gender, or disability~\cite{raghavan2}.
However, in the case of inadvertent discrimination, simply denying a model access to protected attributes does not guarantee an unbiased decision-making process, due to the presence of proxies to protected attributes~\cite{raghavan2}. 
\citet{mehrabi1} denote $23$ different ways that bias can exist in AI systems, more than 5 of which come from bias in the collection and interpretation of input data alone. 
Consequently, unless an AI system has been carefully vetted for the existence of bias, the existence of inevitable bias in the AI pipeline can be expected as the null hypothesis. 
This work is the first to investigate the fare pricing algorithms employed by ridehailing companies for evidence of AI bias.
Results show that fare pricing based on demand relative to supply and trip duration leads to AI bias significantly associated with ethnic makeup, poverty levels, age, and education levels of neighborhoods.

Uber determines where to route their supply of gig workers by predicting demand for future rides using AI models that forecast based on historical demand information~\cite{hermann1}. 
For fare pricing, Uber charges passengers based on the duration and length of their trips, as well as the times that riders use the application~\cite{uber1, uber3}.
In addition, Uber has been known to use the concept of a ``surge multiplier''~\cite{uber1, chen1}, which increases prices depending upon demand relative to supply in a trip's location~\cite{chen1, gerte1}.
Ridehailing company Lyft also has a similar concept to the ``surge multiplier'', called ``primetime pricing''~\cite{banerjee1}.
``Surge multipliers'' vary based on independent ``surge areas'' within cities, which have independent ``surge multipliers''~\cite{chen1}.
Uber pricing varies based upon the location in the city from which rides begin and end. 
An anonymous Uber driver from the city of Chicago stated of the incentive areas, ``low-income areas surged due to these zones'', leading to higher prices in those particular low-income neighborhoods.
Pricing changes caused by incentive zones promoted by ridehailing service providers are troubling, as they may manipulate traffic, supply in neighborhoods, and ride forecasting, possibly motivating disparate impact and misrepresentation in future urban development.

This is the first analysis of AI bias in dynamic pricing algorithms of ridehailing, performed using a dataset of over $100$ million rides from the city of Chicago from November 2018 to September 2019~\cite{ridehailing_data}.
Data from the city of Chicago was chosen because of a new city-wide law requiring ridehailing applications to disclose fare prices~\cite{cnbc1}, making the Chicago ridehailing data the most complete dataset for understanding ridehailing fare pricing and price discrimination models currently available. 
The data includes fare prices and trip duration, as well as pickup and dropoff location, which can be used to connect the dataset to census tracts as designated by the American Community Survey (ACS)~\cite{acs1}. 
The ACS is an annual survey conducted by the federal government of the U.S. that collects demographic statistics about people residing in the U.S. census tracts. 
Census tracts are approximations of neighborhoods, subdivisions of a county designated by the U.S. Census Bureau consisting of about $1,200$-$8,000$ people.
Since ``surge areas'' in ridehailing are not publicly available, fare pricing is examined by location using ACS census tracts~\cite{acs1}.

The study presents a measurement of AI bias in the city of Chicago ridehailing dataset by quantifying associations between fare prices in neighborhoods and their corresponding demographic attributes.
An application of random effects modeling drawn from meta-analysis literature computes city-level AI bias on fare pricing.
This approach provides a thorough and repeatable blueprint for identifying and addressing bias in AI systems that produce outcomes using data from U.S. geolocations. 
Findings reveal that neighborhoods with large non-white populations, higher poverty levels, younger residents, and high education levels correspond to higher trip fares, with combined effect sizes, measured in Cohen's $d$~\cite{cohen1}, of $-0.32$, $-0.28$, $0.69$, and $0.24$ respectively. 
Our results indicate that fare pricing algorithms learning from demand, relative supply, and duration of trips may be the source of disparate impact in ridehailing. Neighborhood residents have no control over demand, supply, or trip duration, but are paying for the consequences of all three.
\section{Related Work}
\label{sec:related_work}

Current anti-discrimination laws do not account for AI bias as it may propagate in a predictive AI model~\cite{barocas2}. Consequently, neither the model nor the ground truth data used to train the model may be available for examination of bias~\cite{barclay1, strobel1, eu1}.
The lack of extensive algorithmic audit and the black box nature of ridehailing fare pricing algorithms leads to the concern of whether they may be unintentionally exhibiting AI bias towards riders based on their demographics or the properties of their neighborhoods.

We examine two different categories of related work in the space of algorithmic bias: research examining bias specifically in ridehailing and taxi service pickup rates~\cite{brown1, ge1, garz2020consumer}, and research detailing how to ensure the fairness of supervised learning models~\cite{dwork1, feldman1, zemel1, cohen2019price, ramandata}.
We also examine two papers on the use of ACS demographic data rather than individual demographic attributes to make inferences about an individual~\cite{geronimus1, soobader1}.

\citet{dillahunt2017uncovering} recruited 13 low-income individuals to study Uber as passengers living in transportation-scarce environments. The analysis revealed barriers such as cost, limited payment methods, and low digital literacy, that make ridehailing services infeasible.
\citet{ge1} examine bias in Uber and Lyft pickups in the cities of Boston and Seattle by requesting human subjects to call for rides using Uber and Lyft, evaluating wait times and cancellation rates based on subject ethnicity through a user study.
Their findings suggest male riders with African-American names are three times more likely to have rides cancelled and wait as much as 35\% longer for rides.
\citet{brown1} performs an extensive analysis of ridehailing applications and their effect on mobility in the city of Los Angeles, including an analysis of the different socioeconomic characteristics of the neighborhoods served by Lyft. \citet{garz2020consumer} seek to bridge regulator and academic perspectives on underlying sources of harm for consumers and potential consequences such as high and hidden prices, post-contract exploitation, and discrimination.
Unlike \citet{ge1}, our work does not rely on a user study, and unlike both \citet{ge1} and \citet{brown1}, our work is the first to provide an analysis on ridehailing trip fare pricing rather than frequency.

\citet{ramandata} studied the impact of rider-driver matching policy and redistributing income to reduce forms of inequality in ridehailing. The analysis revealed that policies minimizing the variance of income also maximize riders, especially during high demand periods. This finding indicates that reducing inequality can boost utility. \citet{cohen2019price} analyze the application of fairness constraints in price discrimination algorithms' price, demand, consumer surplus, and no-purchase valuation.
\citet{feldman1} and \citet{zemel1} detail methodologies for testing and removing bias from datasets, on the same datasets with different perspectives on fairness, disparate impact and statistical parity respectively.
While the debiasing of a supervised model is not relevant to this work directly, these works describe methods for quantifying forms of bias and inequality.
Generally, public access to white box models making decisions algorithmically or the data used to train them is not granted due to privacy regulations~\cite{eu1}, as in prior cases of bias in online recruitment, online advertisements, and recidivism prediction~\cite{datta2015automated, turner_lee1, sweeney2013discrimination}.
The focus of our work is on measuring bias in individualized pricing for riders in neighborhoods defined by ACS. We use observable model outcomes and a partial set of input features, which can be a necessity if models and their training data cannot be made publicly available.

\citet{geronimus1} and \citet{soobader1} test how the use of ACS demographic data to make inferences about individuals compares to the use of individual demographic data to make inferences in the context of predicting health outcomes.
\citet{geronimus1} and \citet{soobader1}'s findings conclude that the results of inferences based upon aggregates can not be interpreted the same as the results of inferences based on individual demographic attributes.
However our methodology does not attempt to make inferences based on neighborhood-wide demographic data, but rather uses bias found across neighborhoods to test bias on individual demographics. 
The role of our methodology for investigating algorithmic decision-makers rather than making further inferences using their outcomes provides the justification for the use of ACS data in our analyses.
In addition, without access to training data, this approach is currently a scientifically justified possible option for analyzing fairness and bias.
\section{Data}

This study mainly focuses on ridehailing data from the city of Chicago and ACS data for Chicago's census tracts. Data from taxis in Chicago are also used to compare ridehailing to taxi fare pricing and coverage, whenever a comparison is possible. The ridehailing and taxi datasets used in this study are collected from the city of Chicago's data portal~\cite{ridehailing_data, taxidata}. Prior to 2019, there was no comparable comprehensive ridehailing dataset available for analysis.

\subsection*{Ridehailing data}
The ridehailing data contains $100,717,117$ trips in the city of Chicago from November of $2018$ to September of $2019$, and the taxi data contains $19,496,933$ ride samples in Chicago during the same time period.
Shared trips, or rides that involve the pickup and dropoff of multiple individuals, are excluded from the ridehailing dataset.
These trips were not included because they contain multiple riders, have multiple pickup and dropoff points as well as a different pricing strategy than rides with only a single rider~\cite{uber2}. 
In addition, including them would limit comparisons of ridehailing services with taxi services, as the taxi services do not list a similar ride sharing option in the provided dataset.
After excluding shared rides and removing any rows with missing data from the dataset, a total of $\sim$68 million ridehailing rides remain.

The dataset contains information about the time, charges, and pickup and dropoff locations for each ride, with location being given in the form of ACS census tracts. 
ACS census tracts are subdivisions of a county designated by the U.S. Census Bureau consisting of about 1,200-8,000 people, whose boundaries are updated every 10 years.

The start and end times of trips are rounded to the nearest 15 minutes and fares are rounded to the nearest $\$2.50$. 
No rider characteristics are reported in order to preserve the privacy of the riders~\cite{chicago2}.
The dataset contains information about the trip duration, charges, and pickup and dropoff locations for each ride, with location being given in the form of ACS census tracts. 
Trips are not labelled by ridehailing service, however, Uber takes roughly $72\%$ of the ridehailing market share in Chicago, and Lyft and Via take up $27\%$ and $1\%$ of the market share~\cite{cnbc1}.

This work is quantifying differences in fare price across neighborhoods defined by census tracts and not on individual riders.
The demographics of individual riders are not available, and an individual rider's demographics cannot be assumed based on the neighborhoods where they are picked up or dropped off.
Instead AI bias measurements, as detailed in the methods section, are taken based on neighborhood demographics.
Other works examining correlations between ridehailing trips and neighborhood demographics~\cite{brown1, lavieri1, gerte1} also perform correlational analysis on ridehailing data and census demographic attributes in a similar fashion.

\subsection*{Additional Preprocessing: Ridehailing data}
Prior to calculating combined effect sizes of bias (detailed in the methods section) on fare price per mile, trip price per mile, and duration per mile are averaged for each census tract, in order to compare census tracts to each other.
Pickup and dropoff counts are summed for each census tract, and then divided by the square meters of each census tract. 
Pickup and dropoff counts are spatially aggregated to account for density of rides in a census tract, rather than raw frequency, as in similar ridehailing trip frequency analyses~\cite{brown1}.

\subsection*{Preprocessing: Taxi data}
Preliminary analysis of the taxi datasets revealed many fare prices inconsistent with the rates charged by Taxi cab companies in Chicago, possibly due to the variety of taxi cab companies in the city and the self-reported nature of the taxi fare pricing data.
For example, Yellow Cab Chicago charges a base fare of $\$3.25$, with an additional charge of $\$2.25$ per mile~\cite{yellowcab1}.
Given that the taxi dataset contains around $13$ million rides when rows with missing data are removed, representative of approximately $37$ million miles worth of driving, assuming all rides were taken using Chicago Yellow Cab, the total expected fare price would be approximately $\$125$ million.
However, the observed overall fare price is closer to $\$143$ million.  
Multiple rides cost much more in fare per mile than expected given listed cab prices, with over $1,000$ rides costing more than $\$1,000$ per mile, and the maximum fare price per mile being approximately $\$90,000$.
Rides with outlier fare prices were detected and removed using a Gaussian Mixture Model (GMM)~\cite{reynolds1}.

GMMs are a probabilistic model which assume data is generated from a mixture of a number of Gaussian distributions, and have been shown to be an effective approach for anomaly detection in prior work~\cite{laxhammar1, li1}.
GMMs require the specification of a number, $k$, of clusters to create.
Taxi rides are clustered assuming $2$ Gaussian distributions, one corresponding to normal rides, and another for anomalous rides, meaning $k$ is set to $2$.
After removing rides clustered in the anomalous cluster, the new total observed fare price is approximately $\$130$ million, down from $\$143$ million, which is closer to the expected total fare of $\$125$ million. 
The new maximum observed fare price per mile is $\$17$, which is more consistent to expectations based on Yellow Cab rates~\cite{yellowcab1} compared to the original maximum fare price per mile of about $\$90,000$.
After removing outliers $10,292,081$ taxi rides remain.

\subsection{American Community Survey Data}
ACS census tract demographic information is retrieved from the United States Census Bureau from the year $2018$.
$2018$ data was used as this is the closest year to the time frame of the Chicago ridehailing dataset for which ACS census data was available.
The average age of ridehailing users as of $2017$ is $37$~\cite{clewlow1}.
In our representation of age information for census tracts, age is split using a threshold of $40$ because age is organized in multi-year brackets in ACS data ($0$-$5$ years old, $5$-$9$ years old, etc), and the age $40$ is the closest bracket age that can be used to divide the population into those above and those below the average ridehailing user age.
A poverty level threshold income of $\$25,000$ is chosen based on the poverty line as defined by United States Department of Health and Human Services~\cite{burwell1}.
The threshold for median house prices is set at $\$250,000$ based on a median house price estimate conducted by Zillow in $2018$~\cite{zillow1}.

All source code and the data analysis performed in this work are available at \url{https://github.com/akshat0123/RShareChicago}.
\section{Methods}
\label{sec:matmeth}

Effect size has been used as a metric to assess implicit human bias in prior works, including the Implicit Association Test~\cite{greenwald1}, which measures differential association of stereotypical attributes and social groups based on nationality, gender, race, religion, and sexuality.
More recently, it has also been used to reveal and quantify biases in unsupervised machine learning approaches that generate representations of language, namely word embeddings~\cite{caliskan2017semantics}.
Cohen's $d$~\cite{cohen1} is a widely-used metric for calculating effect size of bias, which compares the difference of means of effect between two groups, to determine the magnitude of the effect of a treatment.
Cohen's $d$ can be used to measure AI bias because it quantifies the difference in fare pricing between trips in neighborhoods with different demographic statistics that represent distinct social groups.

\subsection{Measuring Effect Size of AI Bias}

An effect size calculation measures the difference in observed outcomes of two populations resulting from a single demographic intervention. 
As an example, given two groups of neighborhoods $A$ and $B$, an effect size of -1 says the mean fare price in $A$ is one standard deviation less than the mean fare price in $B$.
In the context of the demographic attribute non-white percentage, if $A$ includes neighborhoods less than $50\%$ non-white, and $B$ includes all remaining neighborhoods, effect size will quantify the difference in mean fare price between majority white ($A$) and majority non-white neighborhoods ($B$).

Lone effect sizes cannot however, quantify the overall association between a continuous-valued demographic attribute and corresponding observed outcomes.
The standardized mean difference in fare price between majority white and majority non-white neighborhoods only quantifies how fare prices differ when the non-white percentage is $50\%$. 
A complete association measure of fare price and non-white percentage requires that the difference in fare price at every non-white percentage interval for the continuous-valued attribute be accounted for.
As a result, combining multiple effect sizes from different demographic intervals provides a more extensive association measure between a demographic attribute and fare price.

In order to calculate the association between each neighborhood demographic attribute and fare price, a combined effect size is calculated for each demographic attribute using a random effects model~\cite{borenstein1}.
Random effects models from the validated meta-analysis literature summarize effect sizes from across studies, as detailed in Equation~\ref{eq:bias}, by weighting each result with the within-study and between-study variances~\cite{hedges1983random}.
In this work, random effects models are used to combine effect size measurements across demographic intervals to produce a comprehensive measurement of AI bias.
Further information about combined effect sizes are provided in the appendices. 

\begin{figure}[h!]
   \centering 
   \text{Combined Effect Size } ($ces$) \\

\begin{align}
    ces(X, Y) &= \frac{ \sum_{t=[t_{min}...t_{max}]} d(X, Y, t) \times w(X, Y, t) }{ \sum_{t=[t_{min}...t_{max}]} w(X, Y, t) } \label{eq:bias} \\
   \nonumber \\
    d(X, Y, t) &= \frac{ \overline{Y}_{x_{t-}} - \overline{Y}_{x_{t+}} }{ \sigma(Y) }\nonumber\\
  \nonumber \\
    w(X, Y, t) &= \frac{1}{var(X, Y) + var_t(X, Y, t)} \nonumber\\
    \nonumber \\
    var(X, Y) &= \underset{t=[t_{min}...t_{max}]}{\sigma^2}[ d(X, Y, t) ] \nonumber\\
    \nonumber \\
    var_t(X, Y, t) &=  \frac{ |X_{t-}| + |X_{t+}| }{ |X_{t-}| |X_{t+}| } + \frac{ d(X, Y, t) }{ 2 (|X_{t-}| + |X_{t+}| - 2) }  \nonumber\\
    \nonumber\\
    &\times \left( \frac{ |X_{t-}| + |X_{t+}| }{ |X_{t-}| + |X_{t+}| - 2 } \right) \nonumber
\end{align}
\caption{This equation displays how combined effect sizes are calculated.
Random effects models from the validated meta-analysis literature summarize effect sizes from across studies by weighting each result with the within-study and between-study variances~\cite{borenstein1}.
$t$ represents the threshold by which census tracts are being compared.
For example, when comparing neighborhoods by non-white percentage, a threshold of 50 means fare pricing in majority non-white and majority white neighborhoods are being compared. 
$d(X, Y, t)$ corresponds to a calculation of Cohen's $d$~\cite{cohen1} at threshold $t$, and $w(X, Y, t)$ denotes the weight for the Cohen's $d$ calculation at threshold $t$. 
$var(X, Y)$ is the variance across all effect size calculations for a demographic attribute $X$, and $var(X, Y, t)$ is the variance of a single Cohen's $d$ calculation~\cite{coe1}.
}
\label{eq:IESB}
\end{figure}


\subsection{Calculating Combined Effect Sizes}
Combined effect sizes are a measurement of association between a demographic attribute $X$ with a value below a threshold $t$ and target outcomes (fare price), $Y$.
Groups $X_{t-}$ and $X_{t+}$ are subsets of the data where the demographic attribute $X$ is below $t$ ($X_{t-}$) and above $t$ ($X_{t+}$).
Comparing the groups $X_{t-}$ and $X_{t+}$ quantifies how much having an $X$ value below $t$ changes mean output values.
With regard to AI bias in the case of ridehailing, this determines the magnitude of the difference in outcomes (fare pricing) when comparing two sets of neighborhoods.
Examining effect sizes rather than computing correlation coefficients provides the advantage of being able to pinpoint bias at particular thresholds ($t$ values) when necessary, and is also free of any assumption about the linearity or monotonicity of the data, when compared to Pearson~\cite{benesty1} or Spearman~\cite{zar1} correlation. 
In addition, other methods of determining feature importance, like information gain \cite{kullback2}, cannot be applied due to the continuous nature of $X$ and $Y$ in the context of this work. 

\begin{figure*}[h]
  \begin{minipage}[c]{0.64\textwidth}
    \includegraphics[width=1.0\textwidth]{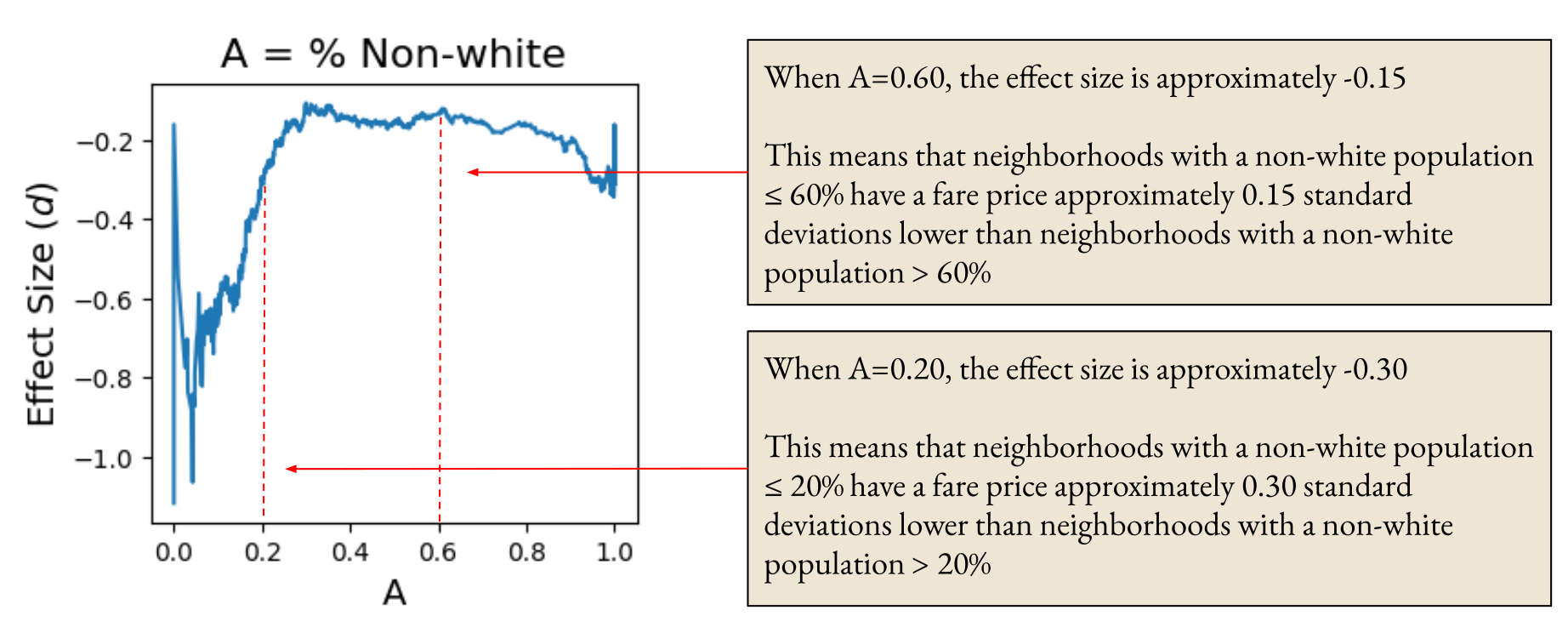}
  \end{minipage}
  \begin{minipage}[c]{0.35\textwidth}
    \vspace{-3mm}
    \caption{This figure displays how effect size charts in this paper can be interpreted.
    The chart displayed shows the effect size of the non-white percentage of a neighborhood's population on ridehailing fare pricing. 
    In this chart, when $A=0.20$, the corresponding effect size is approximately $-0.30$.
    This shows that neighborhoods that are less than $20\%$ non-white have a mean fare price $0.30$ standard deviations lower than neighborhoods that are greater than $20\%$ non-white.
    }
    \label{fig:read_es}
    \end{minipage}
\end{figure*}

Effect sizes calculated between groups $X_{t-}$ and $X_{t+}$ as $t$ is increased from the minimum to the maximum observed values of $X$ are combined using a weighted mean of effect sizes, where weights are assigned using a random effects model~\cite{borenstein1}. Figure~\ref{fig:read_es} goes over an example of measuring effect size of the proportion of non-white residents in a neighborhood on fare pricing along with information for interpreting the chart's $x$ and $y$ axes values.

In meta-analysis, random effects models are used to compute a combined effect given a set of effect size calculations, where weights are assigned to effect sizes given the amount of information they contain~\cite{borenstein1}.  
A random effects model is used to summarize the effect sizes rather than computing a mean of effect sizes because as $t$ increases, the sizes of groups $X_{t-}$ and $X_{t+}$ change, which effects the variance of each effect size calculation.  A random effects model is also called a variance components model where the model parameters are random variables.
Fixed effects model does not apply in this scenario since the group means are not fixed, as opposed to a random effects model in which the group means are a random sample from a population.

In equation~\ref{eq:bias}, $d(X, Y, t)$ corresponds to a calculation of Cohen's $d$~\cite{cohen1} at threshold $t$, and $w(X, Y, t)$ denotes the weight for the Cohen's $d$ calculation at threshold $t$. 
$var(X, Y)$ is the variance across all effect size calculations for a demographic attribute $X$, and $var(X, Y, t)$ is the variance of a single Cohen's $d$ calculation~\cite{coe1}.

The statistical significance ($p$-value) for combined effect sizes are calculated using permutation testing~\cite{ptest}, which measures the likelihood of the null hypothesis, by computing the probability that randomly selected groups of census tracts (with any $t$ setting, from Equation~\ref{eq:bias}) could produce a larger output than a combined effect size calculation.
The permutation test is $Pr_i[ces(X_i, Y_i) > ces(X, Y)]$, where $X_i$ and $Y_i$ are a randomly selected subset of $X$ and $Y$, which account for census-wide data for 1,159 census tracts for the ridehailing dataset and 820 census tracts for the taxi dataset.
1,000 iterations of the permutation test are run to calculate p-values for each reported AI bias measurement.

\subsection{Interpreting Effect Size Charts}
Figure~\ref{fig:read_es} provides a guideline on how effect size charts can be interpreted. 
In Figure~\ref{fig:read_es}, when $A=0.20$, the corresponding effect size is approximately $-0.30$.
This shows that neighborhoods that have less than $20\%$ non-white residents have a mean fare price $0.30$ standard deviations lower than neighborhoods that have more than $20\%$ non-white residents.
Every setting of $A$ in the chart can be read similarly.
When $A=0.60$, the corresponding effect size is approximately $-0.15$.
This shows that neighborhoods that have less than $60\%$ non-white residents have a mean fare price $0.15$ standard deviations lower than neighborhoods that have more than $60\%$ non-white residents.
Using charts such as these, the neighborhoods with the large and small differences in mean fare pricing can be distinguished.
In Figure~\ref{fig:read_es}, effect size is lowest approximately when $A$, or non-white percentage is approximately $5\%$. 
This shows that neighborhoods with less than a $5\%$ non-white population have the largest difference in mean fare price, approximately $-1.10$ standard deviations away from fare prices in neighborhoods with greater than a $5\%$ non-white population.

\section{Results}

\begin{figure*}[h]
\centering
    \includegraphics[width=\textwidth]{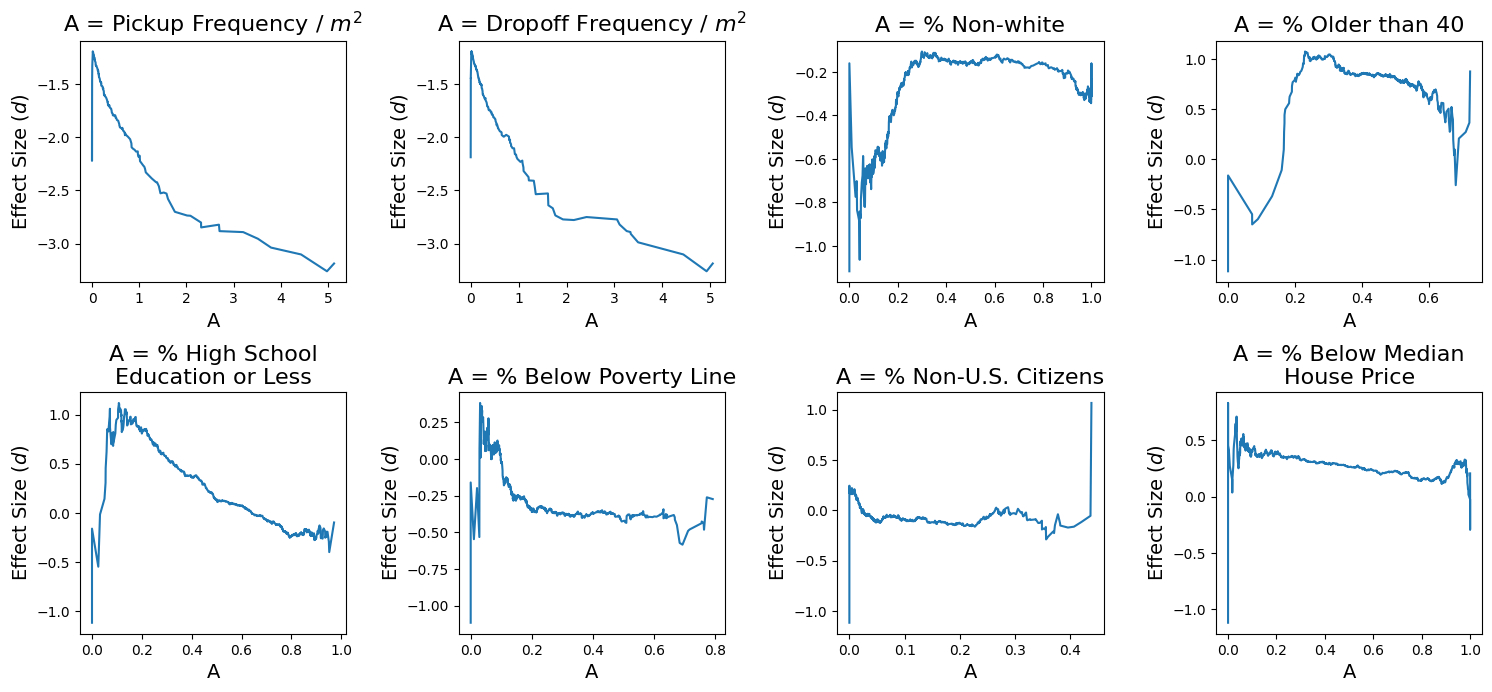}
    \caption{Effect sizes on fare price per mile are calculated for pickup and dropoff frequencies per neighborhood. 
    Effect sizes on fare price per mile are also calculated for the percentage of non-white, over 40, high-school educated or less, below poverty line, and non-U.S. citizen populations, as well as the percentage of people living in homes below median home value in Chicago.
    Effect sizes are calculated using Cohen's $d$~\cite{cohen1}.
    As an example, in the chart labelled ``A = \% Non-White'', when $x=0.2$, the effect-size is $y=-0.6$ ($d$). 
    This says the average fare price per mile for neighborhoods less than 20\% non-white is 0.6 standard deviations less than the average fare price per mile for neighborhoods more than 20\% non-white.
    Accordingly, the more the non-white populations in a census tract, the more expensive the fare price. 
    }
    \label{fig:price_effect_size}
\end{figure*}

In relation to ridehailing, AI bias measures a difference in variables of the dataset based on neighborhood demography.
For fare pricing, this difference in outcomes is measured as the difference in mean fare price between neighborhoods with different demographic characteristics. 
The standardized difference in mean fare price is Cohen's $d$~\cite{cohen1}, or the effect size.
Conventional small, medium, and large values of Cohen's $d$ are $0.20$, $0.50$, and $0.80$.
In Figure~\ref{fig:price_effect_size}, effect size of AI bias measures the differential fare pricing associated with two groups of neighborhoods with different ratios of a demographic attribute. For example, the chart with the $x$-axis A = \% Non-U.S. Citizens in Figure~\ref{fig:price_effect_size}, measures the effect of the variable percentage of immigrants in a neighborhood, on fare pricing.

AI bias effect sizes, summarized using random effects modeling, show that fares, trip frequency, and trip duration are all lower for trips beginning or ending in neighborhoods with older populations (see Tables~\ref{table:price_table} and~\ref{table:freq_time_table}).
Trips beginning in neighborhoods with a smaller ratio of educated residents, as well as neighborhoods with lower home values also have lower fares, trip frequency, and trip durations (see Tables~\ref{table:price_table} and~\ref{table:freq_time_table}).
Trips beginning or ending in neighborhoods with more non-white residents, or more residents living below the poverty line, have higher fares, as well as higher trip duration (see Table~\ref{table:price_table}).
However, in both of these groups of neighborhoods (non-white, high level of poverty), demand is lower (see Table~\ref{table:freq_time_table}).
The percentage of immigrants in a neighborhood does not show any significant association with fare pricing (see Table~\ref{table:price_table}).

Pickup and dropoff frequencies have a strong positive association with fare pricing, meaning that neighborhoods with more frequent trips have higher prices (see Table~\ref{table:price_table}).
The positive relationship between pickup and dropoff frequencies and fare price is consistent with Uber's ``surge pricing'' strategy~\cite{uber5} - areas where there are more trips or demand are more likely to have relatively fewer drivers, or supply, and go into ``surge'', causing prices to increase.
This pattern is not observed in neighborhoods with more non-white residents, or with more residents living below the poverty line, which is an anomaly. However, if there is very low supply of drivers in these neighborhoods, demand relative to the supply and accordingly fare pricing might increase. Ridehailing providers do not publish data regarding where and why they route the supply of drivers and which areas they declare as unsafe no go zones. As a result, analyzing AI bias caused by algorithm's surge pricing objective that relies on supply is currently not possible. Having access to this information would make it possible to understand how algorithmic supply routing may affect the traffic conditions, relative demand, and the corresponding fare pricing.

\begin{figure*}[h]
    \centering
    \includegraphics[width=\textwidth]{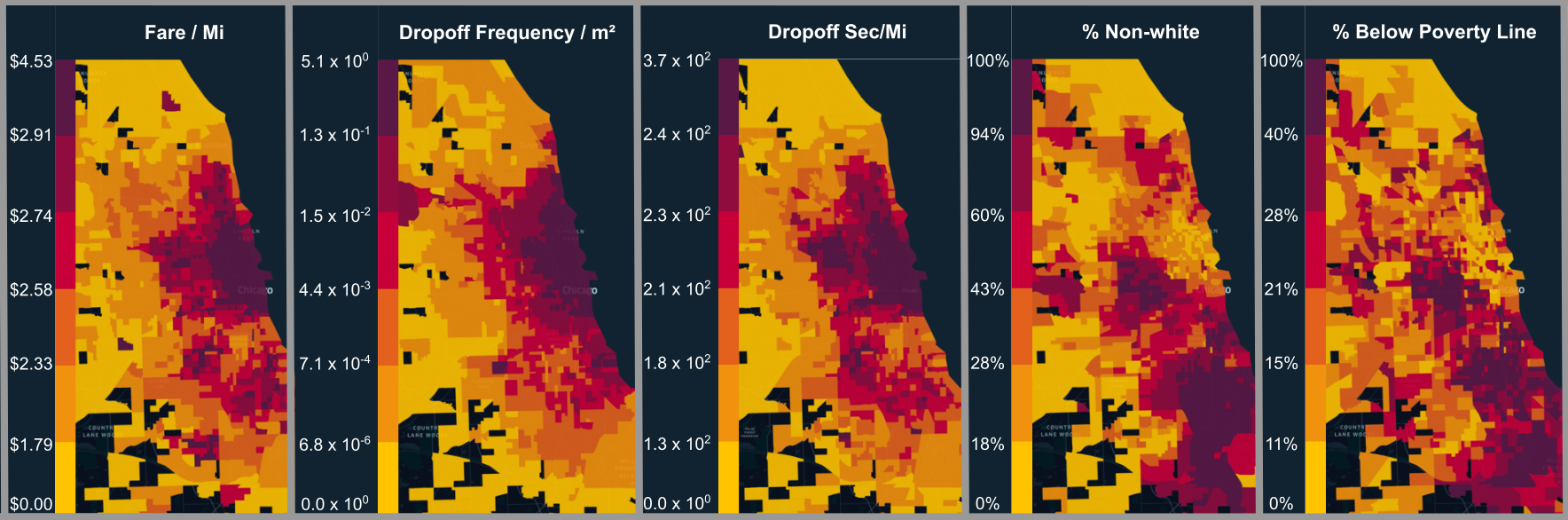}
    \caption{
    The heat maps depict, in order, fare pricing, ride frequency, ride duration, ethnic makeup, and poverty levels across the city of Chicago.
    The fare pricing map shows average fare per mile prices in a census tract in dollars, the dropoff count map shows the average number of dropoffs per square meter in each census tract, and the ride duration map shows the average number of seconds per mile for trips in a census tract.
    The remaining maps show the amount of each demographic labelled in their respective titles by percentage in each census tract.
    The color coding for each map is scaled by 6-quintiles, with darker colors representing higher values and lighter colors representing lower values for each respective neighborhood attribute.
    Non-white neighborhoods and neighborhoods with a large population living below the poverty line have higher fare prices (combined effect size $d=-0.32$, $-0.28$). 
    Non-white and impoverished neighborhoods also show lower demand (combined effect size $d=0.25$, $0.27$), but increased trip duration (combined effect size $d=-0.31$, $-0.25$). Increased fare prices in these areas are anomalous since low demand can cancel increased duration's effect in fare pricing. Consequently, ridehailing algorithms' notion of surge pricing based on relative supply and demand might be the underlying source of bias leading to disparate impact.
    }
    \label{fig:heatmaps}
\end{figure*}

The highest demand neighborhoods for ridehailing trips in Chicago are in the city's business district, known as ``The Loop'' (see Figure~\ref{fig:heatmaps}).
Neighborhoods with a high percentage of non-white residents, and neighborhoods where much of the population live below the poverty line, lie south of Chicago's high-demand business district (see Figure~\ref{fig:heatmaps}).
Most of these neighborhoods (non-white, high poverty level) are concentrated in the southwest corner of Chicago, where trip durations are low, while others are just southwest of the business district, where trip durations are high (see Figure~\ref{fig:heatmaps}). 

Combined effect size analysis for a taxi fare dataset from Chicago collected from the same time period as the ridehailing data shows that the demographic attributes of neighborhoods where taxi trips begin have no significant association with fare pricing (see Table~\ref{table:price_table}). 
Taxi rides that end in neighborhoods with older populations, less educated residents, or lower value homes have cheaper fare prices (see Table~\ref{table:price_table}).
The association between taxi fare pricing and age of the population is weaker than the association between ridehailing fare pricing and age of the population.
Taxi fare pricing has a stronger association with both college education levels and home values in a trip destination neighborhood than ridehailing fare pricing does.

\begin{table*}[ht]
    \centering
      \begin{adjustbox}{max width=\textwidth}

        \begin{tabular}{l|c|c|c|c|c|c|c|c}
            & \multicolumn{4}{|c|}{\textbf{Ridehailing Fare Price/Mile}} & \multicolumn{4}{|c}{\textbf{Taxi Fare Price/Mile}} \\\hline
            & \multicolumn{2}{|c|}{\textbf{Pickup}} & \multicolumn{2}{|c|}{\textbf{Dropoff}} & \multicolumn{2}{|c|}{\textbf{Pickup}} & \multicolumn{2}{|c}{\textbf{Dropoff}} \\\hline
            \textbf{Attribute} & \textbf{Combined} & \textbf{p} & \textbf{Combined} & \textbf{p} & \textbf{Combined} & \textbf{p} & \textbf{Combined} & \textbf{p} \\
            & \textbf{Effect Size (d)} & & \textbf{Effect Size (d)} & & \textbf{Effect Size (d)} & & \textbf{Effect Size (d)} & \\\hline
            Pickup Frequency / $m^2$                     & -1.57 & $<10^{-3}$ & -1.59 & $<10^{-3}$ &  0.24 & 0.04 & -0.96 & $<10^{-3}$ \\
            Dropoff Frequency / $m^2$                    & -1.57 & $<10^{-3}$ & -1.57 & $<10^{-3}$ & -0.15 & 0.21 & -0.88 & $<10^{-3}$ \\
            \% Non-white                              & -0.22 & 0.02       & -0.32 & $<10^{-3}$ &  0.18 & 0.11 &  0.07 & 0.50       \\
            \% Older than 40                          &  0.66 & $<10^{-3}$ &  0.69 & $<10^{-3}$ & -0.04 & 0.72 &  0.52 & $<10^{-3}$ \\
            \% High School education or less            &  0.24 & 0.01       &  0.15 & 0.17       &  0.05 & 0.65 &  0.41 & $<10^{-3}$ \\
            \% Below Poverty Line                     & -0.19 & 0.05       & -0.28 & $<10^{-3}$ &  0.01 & 0.94 &  0.05 & 0.59       \\
            \% Non-U.S. Citizens                      & -0.10 & 0.35       & -0.07 & 0.54       & -0.02 & 0.83 &  0.04 & 0.71       \\
            \% Below Median House Price &  0.23 & 0.02       &  0.19 & 0.06       &  0.02 & 0.85 &  0.32 & $<10^{-3}$ \\            
        \end{tabular}
      \end{adjustbox}

    \caption{Combined effect sizes on fare price per mile by neighborhood attributes - Combined effect sizes scores are shown for the fare price per mile given a set of neighborhood attributes. 
    Combined effect sizes are weighted using a random-effects model~\cite{borenstein1}.
    The ``Ridehailing'' column contains combined effect sizes calculated for ridehailing trip fares, and the ``Taxi'' column contains combined effect sizes calculated for ``Taxi'' trip fares.
    ``Pickup'' and ``Dropoff'' columns designate fare price per mile when being picked up or dropped of in a neighborhood and ``p'' presents the p-value for effect size calculations.}
    \label{table:price_table}
\end{table*}

\begin{table*}[hbt!]
    \centering
          \begin{adjustbox}{max width=\textwidth}

    \begin{tabular}{l|c|c|c|c|c|c|c|c}
        & \multicolumn{4}{|c}{\textbf{Ridehailing Trip Frequency / $m^2$}} & \multicolumn{4}{|c}{\textbf{Ridehailing Trip Seconds/Mile}} \\\hline
        & \multicolumn{2}{|c}{\textbf{Pickup}} & \multicolumn{2}{|c}{\textbf{Dropoff}} & \multicolumn{2}{|c}{\textbf{Pickup}} & \multicolumn{2}{|c}{\textbf{Dropoff}} \\\hline
        \textbf{Attribute} & \textbf{Combined} & \textbf{p} & \textbf{Combined} & \textbf{p} & \textbf{Combined} & \textbf{p} & \textbf{Combined} & \textbf{p} \\
        & \textbf{Effect Size (d)} & & \textbf{Effect Size (d)} &  & \textbf{Effect Size (d)} & & \textbf{Effect Size (d)} & \\\hline
        \% Non-white                     &  0.25 & $<10^{-3}$ &  0.23 & $<10^{-3}$ & -0.14 & 0.17       & -0.31 & $<10^{-3}$ \\
        \% Older than 40                 &  0.38 & $<10^{-3}$ &  0.38 & $<10^{-3}$ &  0.67 & $<10^{-3}$ &  0.70 & $<10^{-3}$ \\
        \% High School Education or less &  0.69 & $<10^{-3}$ &  0.69 & $<10^{-3}$ &  0.20 & 0.05       &  0.08 & 0.43       \\
        \% Below Poverty Line            &  0.27 & $<10^{-3}$ &  0.26 & $<10^{-3}$ & -0.13 & 0.19       & -0.25 & $<10^{-3}$ \\
        \% Non-U.S. Citizens             & -0.09 & 0.21       & -0.11 & 0.13       & -0.21 & 0.02       & -0.11 & 0.32       \\
        \% Below Median House Price      &  0.42 & $<10^{-3}$ &  0.43 & $<10^{-3}$ &  0.30 & $<10^{-3}$ &  0.17 & 0.08       \\
    \end{tabular}
    \end{adjustbox}
    \caption{
        Combined effect sizes on trip frequency and trip seconds per mile by neighborhood attributes - combined effect sizes are shown given a set of neighborhood attributes. 
        The ``Ridehailing Trip Frequency / $m^2$'' column contains combined effect sizes of neighborhood demographic attributes on the frequency of trips in a neighborhood, and the ``Ridehailing Trip Seconds/Mile'' column contains combined effect sizes measuring the effect size of neighborhood attributes on the time taken per mile for trips in a neighborhood.
        ``Pickup'' and ``Dropoff'' columns designate frequency and trip time when being picked up or dropped of in a neighborhood and ``p'' presents the p-value for effect size calculations.
    }
    \label{table:freq_time_table}
\end{table*}

%
%
%
\section{Discussion}

Prior works have shown that younger people and college educated individuals use ridehailing services more frequently than older people and those without a college education~\cite{brown1, clewlow1}, which is reflected in the measurement of AI bias fare pricing as well as demand measured by trip frequency (see Table~\ref{table:price_table} and Table~\ref{table:freq_time_table}).
In addition, for Chicago's census tracts, the percentage of population older than 40 or without a college degree in a neighborhood is negatively correlated with dropoff frequency (Pearson's $r=-.23, p<10^{-4}$, $r=-.40, p<10^{-4}$).
Higher levels of demand combined with longer trip durations (see Tables~\ref{table:price_table} and Table~\ref{table:freq_time_table}) may explain why younger and highly college educated neighborhoods have higher fare prices.

The difference in fare pricing for neighborhoods with more non-white residents or more residents living below the poverty line could be occurring for a variety of reasons. These neighborhoods are associated with longer trip duration but also decreased demand, which doesn't directly explain the increased fare pricing. One potential reason for decreased demand is the low number of businesses and entertainment options in non-white and impoverished neighborhoods relative to the rest of the city. 
The non-white percentage and poverty level of a neighborhood are both negatively correlated with the number of jobs in arts, food, and entertainment (Pearson's $r=-.13, p<10^{-4}, r=-.11, p<10^{-3}$) as well as the total number of businesses in a neighborhood (Pearson's $r=-.27, p<10^{-4}, r=-.25, p<10^{-4}$). However, the number of ridehailing trips in a neighborhood is positively correlated with the number of businesses and jobs in arts, food, and entertainment (Pearson's $r=.84, p<10^{-4}, r=.88, p<10^{-4}$).

Prior works~\cite{clewlow1, rayle1} have found that the most common reason for using ridehailing services is for leisure activities, such as going to bars and restaurants.
Residents of neighborhoods that are non-white or have high poverty levels may have to travel further to business district neighborhoods in order to pursue the same leisure activities as neighborhoods with lower non-white or below poverty level populations.
Demand and fare pricing in neighborhoods with more businesses and entertainment is higher than in other areas of the city (see Figure~\ref{fig:heatmaps}).
In addition, the non-white percentage and poverty level of neighborhoods are both positively correlated with the number of workers in arts, food, and entertainment (Pearson's $r=.21, p<10^{-4}, r=.34, p<10^{-4}$). 
Many of these workers may also be traveling to neighborhoods where there are more arts and entertainment options for their employment. 
Overall, traveling to or from high demand neighborhoods during rush hour might increase fare pricing for trips originating or ending in neighborhoods with larger non-white or below poverty level populations.

Combined effect size analysis indicates that trips for neighborhoods with more non-white and below poverty level residents have longer trip times (see trip seconds/mile columns in Table~\ref{table:freq_time_table}). 
Increased trip times may be related to trips to higher traffic areas of the city with more businesses and entertainment options.
The number of jobs in arts, food, and entertainment have a positive correlation with traffic and car accidents (Pearson's $r=0.40, p<10^{-4}, r=0.59, p<10^{-4}$).
Similarly, the number of businesses in a neighborhood positively correlates with both traffic and car accidents (Pearson's $r=0.58, p<10^{-4}, r=0.81, p<10^{-4}$).


One potential reason for increased fare pricing in non-white and high poverty level neighborhoods is being subject to surge pricing due to low supply of drivers if the supply of drivers are not distributed uniformly, relative to the demand distribution, around the city. If drivers avoid these neighborhoods and the demand is relatively higher than supply, the ridehailing algorithm's dynamic pricing objective based on the surge principle will perpetuate social bias by increasing the algorithmic fare price. Since the data and corresponding algorithm for ridehailing supply allocation are not available to the public, we are not able to measure the magnitude of bias associated with surge pricing that depends on supply information.

We discuss in detail how another possible reason for high fares in neighborhoods with more non-white residents is driver prejudice and fear of neighborhoods with high crime rates.
Ge et al.~\cite{ge1} found that ridehailing users in Boston with African-American names had rides cancelled more than twice as often as those with white sounding names. 
The same study also found that in Seattle, riders with African-American names had almost a 35\% increase in waiting times. If drivers or the supply allocation algorithm are avoiding non-white and high poverty level neighborhoods, this will cause an increase in fare pricing due to relatively low supply.

Uber drivers have also been reported to avoid high-crime areas on multiple occasions. So called ``no-go zones'', which are areas avoided by Uber drivers, have been identified in Rio de Janeiro~\cite{nogo_brazil}, Johannesburg~\cite{nogo_johannesburg}, and Atlanta~\cite{nogo_atlanta}. 
Uber drivers avoid these areas due to fear of becoming victim of a crime. In 2017, Uber tested a new feature allowing drivers to filter out areas of cities in which they did not wish to drive~\cite{hawkins1} and currently, Uber still allows drivers to set a destination twice a day \cite{uber6}, which matches them to riders going toward their specified destination.

The  percentage of non-white residents in a neighborhood and percentage of below poverty level residents of a neighborhood in Chicago are both positively correlated with crime rates (Pearson's $r=0.23, p<10^{-4}, r=0.27, p<10^{-4}$).
Crime rates are positively correlated with fare prices at both trip pickup location (Pearson's $r=0.47, p<10^{-4}$) and fare prices at trip dropoff location (Pearson's $r=0.49, p<10^{-4}$) as well.
Allowing drivers to choose destinations, as well as driver preferences, in conjunction with demand and relative supply based pricing could be causing prices to rise in areas of high crime due to a lack of driver supply.
A comprehensive safety report released by Uber~\cite{uber7} found a total 7 cases of fatal physical assault against drivers out of 1.3 billion rides taken in the United States in 2018. 
Ridehailing companies should be concerned about algorithmic disparate impact if neighborhoods are being avoided due to fear of crime, direct driver prejudice, or the supply allocation algorithm.
14\% of the rides in Chicago examined in this study are in majority non-white neighborhoods. 

The ridehailing data provided by the city of Chicago is a list of rides, and not a list of ride requests, which are not public. 
Since ride requests are not available, trip frequency in a neighborhood only partially accounts for demand.
Ridehailing companies have detailed information about demand and how the supply of gig workers is routed around the city based on monetary incentives that is not publicly available.

\begin{figure}[t] \begin{center}
    \includegraphics[width=0.35\textwidth]{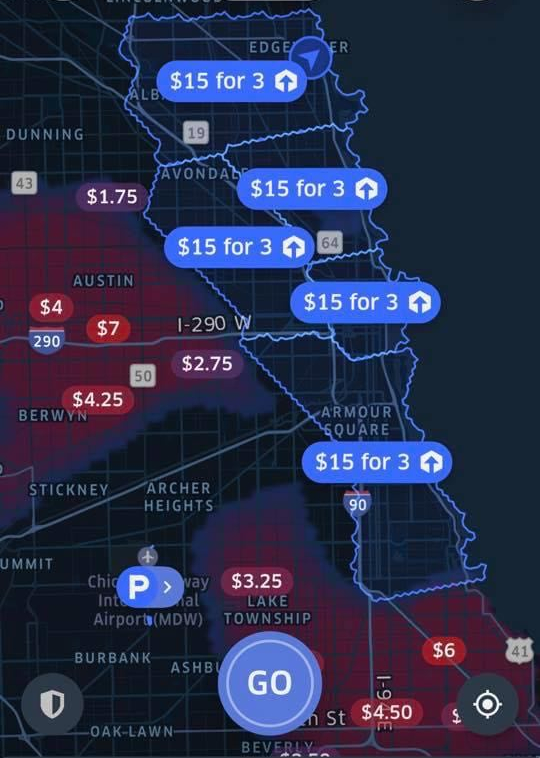} 
    \caption{This image displays an anonymous driver's view when driving for Uber.
    Blue areas with bonuses listed on them (e.g. ``15 for 3'') are boost zones, where drivers are incentivized to receive bonuses when they complete rides.
    Red areas are in surge, meaning that low supply of drivers in these areas resulted in relatively high demand, which in turn caused prices to rise in the area.
    } \label{fig:gigwork}
  \end{center}
\end{figure}

Uber and Lyft both offer bonuses to drivers to drive in particular areas of cities.
Uber offers drivers the ability to earn bonuses by taking rides in ``boost zones''~\cite{uber4} and Lyft similarly offers ``personal power zones''~\cite{lyft1}.
Figure~\ref{fig:gigwork}, provided by an anonymous gig worker, displays how ``boost zones'' and ``surge areas'' work from the perspective of a driver.
Blue areas with bonuses listed on them (e.g.``15 for 3'') are boost zones, where drivers receive bonuses when they complete rides.
Red areas are in surge, meaning that high demand compared to available supply has caused prices in the area to rise.
The driver that provided the screenshot claimed ``low-income areas surged after bonuses were introduced by the ridehailing platform in blue zones". 
The area to the left of the boost zones, ``Austin'', is a community area in Chicago that contains a population that is $95.00\%$ non-white, $55.50\%$ high school educated or less, and $39.20\%$ below the poverty line~\cite{cmap}.
Artificial manipulation of supply by ridehailing companies, incentivizing drivers to shift to other parts of the city, could also be a cause for increased prices by artificially inflating demand in neighborhoods with high percentages of non-white and below poverty level populations.

The main findings of this analysis of Chicago ridehailing data are that fare pricing is higher for neighborhoods with younger, highly college educated, non-white, and high below poverty level populations.
The difference in pricing for younger and highly college educated neighborhoods may be explained by an analysis of demand and trip duration (see Table~\ref{table:freq_time_table}).
The differences in pricing for neighborhoods with more non-white and below poverty level residents may be explained by trip duration, but can not be explained by low demand (see Table~\ref{table:freq_time_table} and Figure~\ref{fig:heatmaps}).
\citet{raghavan2} point out that even in the case when a business argues disparate impact caused by an algorithmic decision maker is justified due to it being the result of a necessary business purpose, the business may still be held liable if an alternative decision making process could perform the same purpose without disparate impact. 
For the case of algorithmic hiring, \citet{raghavan2} suggest that vendors of such models should, at minimum, know the protected classes of individuals during training to prevent disparate impact.
For the case of ridehailing, in order to prevent disparate impact, ridehailing companies should at the very least consider the demography, effect of prejudice and algorithmic allocation on supply, transit geography, and economic geography of the cities in which they operate when designing fare pricing algorithms. Moreover, measuring bias, simulating fairness, or \citet{ramandata}'s approaches to eliminate inequity while increasing utility might help ridehailing providers design a more equitable price discrimination algorithm.

To extend this analysis, obtaining the supply allocation algorithm, as well as data regarding the total number of ride requests and supply in each neighborhood would allow for a more comprehensive understanding of the relationship between demand and fare pricing.
Current analysis is dependent upon trip frequencies in neighborhoods, which does not exhaustively account for relative rider demand.
Information about what rides are declined, and how price changes for a rider after their initial request is declined may further explain why prices are high in areas where there is relatively lower demand.
Understanding how fare prices change in neighborhoods when driver incentive zones are activated can also be used to explain why higher fare prices occur in neighborhoods with relatively low demand.
Explicit detail of how, when, and where an incentive zone is introduced, and how these zones impact trip frequency would also shed light on how ridehailing companies may be inadvertantly manipulating demand and relative supply, and as a result, fare prices.

\section{Conclusion}

This study demonstrates that dynamic pricing algorithms of ridehailing learn AI bias via the ridehailing utilization patterns in neighborhoods with diverse demographic makeup. 
The methodology used in measuring AI bias in ridehailing can be adapted to examine bias in datasets of geolocation based predictive AI models that have continuous outcomes such as pricing. 

According to the publicly available data, ridehailing fare pricing algorithms that partially use trip frequency and duration information in decision making have generated a pricing scheme that results in disparate impact given neighborhood demographics in the city of Chicago.
From the time period between November $2018$ and September $2019$, neighborhoods with younger, more college educated, poorer, and more non-white residents faced higher fares using ridehailing applications in the city of Chicago.
Moreover, current and past pricing may affect future demand, which may in turn change pricing. 
Several iterations of this feedback cycle may cause bias to propagate from fare pricing to service, as the supply of drivers is continuously shifted to higher demand areas.

In addition to existing prejudice, if the social demography, transit geography, and economic geography of a city are not taken into account in the development of dynamic pricing algorithms, there may be disparate impact in pricing outcomes.
If disparate impact in outcomes is not accounted for, it will remain in the data when it is re-purposed for smart city development, where it may cause further disparate impact depending on how the data is used.
From a policy perspective, lack of appropriate transparency measures regarding data collection and AI decision making enforced upon companies may be contributing to disparate impact. 
Ultimately, algorithmic bias cannot be mitigated without algorithmic transparency. 

\section{ACKNOWLEDGEMENTS}
We want to thank Robert Pless and the anonymous reviewers for their useful feedback.

\bibliographystyle{ACM-Reference-Format}
\bibliography{00_main_cameraReady_long}

\newpage

\section{APPENDICES}

\subsection{Calculating Combined Effect sizes}
Equation~\ref{eq:bias} displays how combined effect sizes are calculated using a random effects approach.
Taking the demographic neighborhood attribute ``\% Below Poverty Line'' as an example, the combined effect size is a summary of the corresponding chart in Figure~\ref{fig:price_bias_pickup} (row 2 column 2).
When $t=30$, $d(X, Y, t)$ calculates the effect size on fare price of being picked up in a neighborhood with 30\% of the population below the poverty line versus being picked up in a neighborhood with 70\% of the population above the poverty line (see Equation~\ref{eq:bias}).
$X_{t-}$ are the neighborhoods with a percentage of the population below the poverty line less than $t$, and $X_{t+}$ are the neighborhoods with a percentage of the population below the poverty line greater than $t$.
The combined effect size summarizes the effect sizes calculated between groups $X_{t-}$ and $X_{t+}$ on fare price $Y$ as $t$ is increased from the minimum to the maximum observed values of $X$.

\subsection*{Additional Datasets: Businesses, Traffic, Crashes, Transit, Groceries, Crime, and Arts/Entertainment in Chicago}
Datasets regarding business, average daily traffic, crashes, subway stations, grocery stores, and crimes in Chicago were collected from the Chicago Data Portal~\cite{dataportal}.
All data in these datasets were used to count frequencies of businesses, cars, crashes, subway stations, grocery stores, and crimes in each census tract where corresponding data was available.
No preprocessing steps were performed on these datasets.

The business license dataset contains $54,734$ businesses across $824$ census tracts in Chicago.
The traffic dataset is a collection of $1,279$ average daily traffic counts for the number of vehicles on streets at select locations in Chicago.
The traffic dataset includes average daily counts for locations across $562$ census tracts.
The crash dataset lists information about each traffic crash to occur on city streets within Chicago, comprised of data for $429,415$ crashes across $843$ census tracts.
The transit dataset contains a list of subway stops on Chicago's ``L'' metro system, consisting of $300$ stops across $112$ census tracts.
The grocery store dataset contains a list of grocery stores in Chicago, used to estimate how many Chicagoans live in food deserts, and lists $506$ grocery stores across $342$ census tracts.

The crimes dataset contains a list of all crimes that were recorded in Chicago from November $2018$ to September $2019$, the same time period as the trips in the ridehailing dataset.
The crimes dataset consists of $237,426$ crimes across $838$ census tracts. 
The number of crimes in a neighborhood is positively correlated with fare prices in neighborhoods.
The correlation between number of crimes and fare price, calculated using Pearson's $r$ is $0.47$, with a p-value less than $10^{-4}$, when examining neighborhoods where trips begin.
The correlation between number of crimes and fare price, calculated using Pearson's $r$ is $0.49$, with a p-value less than $10^{-4}$, when examining neighborhoods where trips end.

Data regarding jobs and workers in the arts, food, and entertainment was collected from the Longitudinal Employer-Household Dynamics (LEHD) program of the U.S. Census Bureau~\cite{lehd1}.
The number of jobs in the arts, food, and entertainment as well as the number of workers in the arts, food, and entertainment were collected using the LEHD variables NAICS $71$ (Arts, Entertainment, and Recreation) and NAICS $72$ (Accommodation and Food Services).
The LEHD data contains counts for jobs and workers in arts, food, and entertainment for $1,314$ census tracts in and around Chicago.

\subsection*{Extended Results: Fare Pricing}
\begin{figure*}
    \centering
    \includegraphics[width=0.99\textwidth]{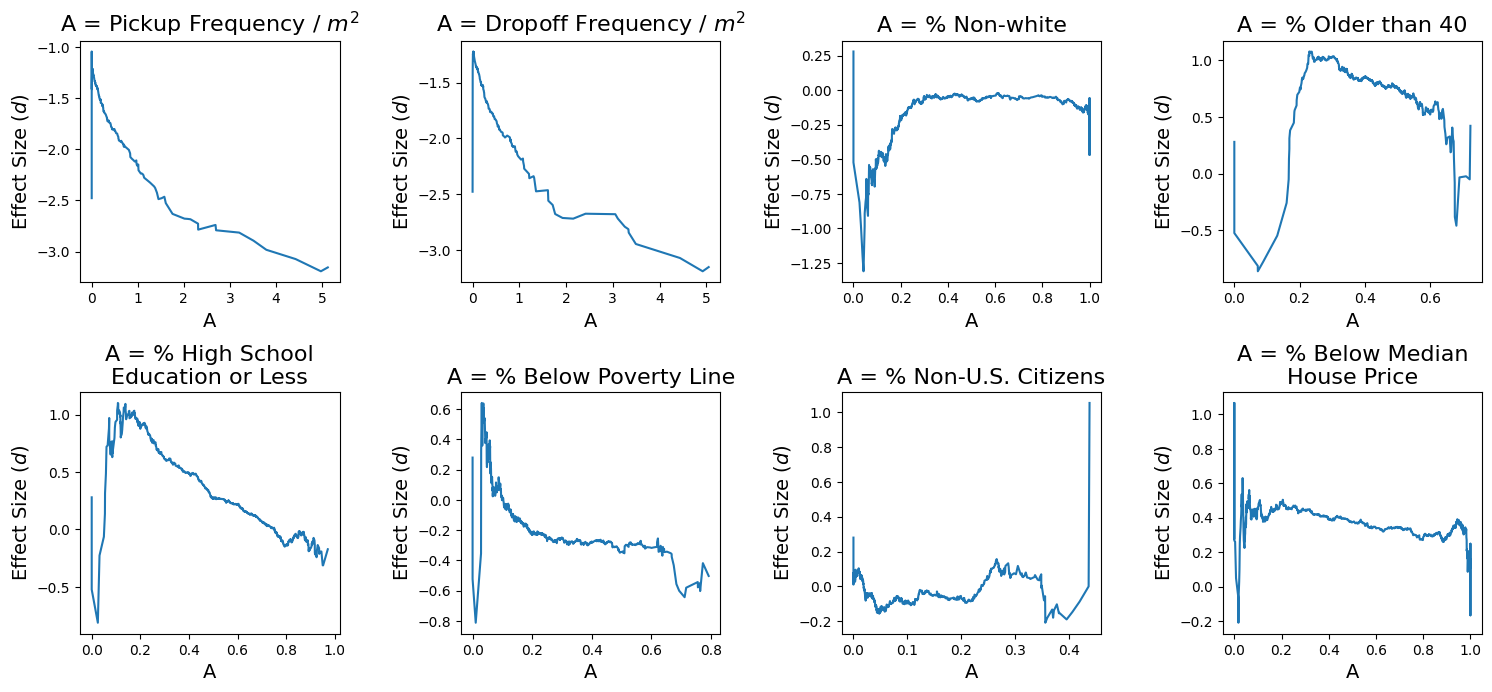}
    \caption{Effect sizes on fare price per mile at pickup location - Effect sizes on fare price per mile are calculated for pickup and dropoff frequencies per neighborhood. 
    Effect sizes on fare price per mile are also calculated for the percentage of non-white, over $40$, high school educated or less, below poverty line, and non-U.S. citizen populations, as well as the percentage of people living in homes below median home value in Chicago.
    Effect sizes are calculated using Cohen's $d$~\cite{cohen1}.
    \textbf{The raw fare prices generating these effect size charts can be observed in detail in Figure~\ref{fig:price_raw_pickup}.} }
    \label{fig:price_bias_pickup}
\end{figure*}

\begin{figure*}
    \includegraphics[width=\textwidth]{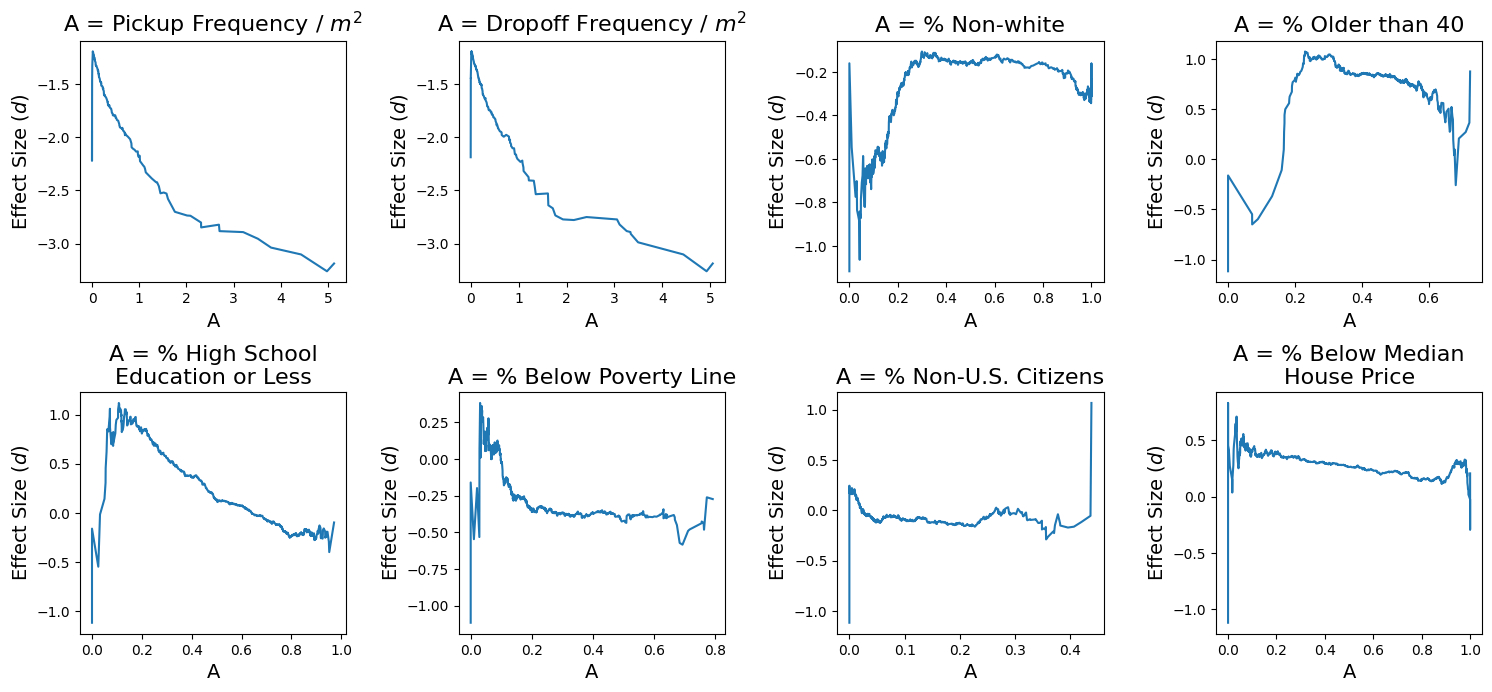}
    \caption{Effect sizes on fare price per mile at dropoff location - Effect sizes on fare price per mile are calculated for pickup and dropoff frequencies per neighborhood. 
    Effect sizes on fare price per mile are also calculated for the percentage of non-white, over $40$, high school educated or less, below poverty line, and non-U.S. citizen populations, as well as the percentage of people living in homes below median home value in Chicago.
    Effect sizes are calculated using Cohen's $d$~\cite{cohen1}.
    \textbf{The raw fare prices generating these effect size charts can be observed in detail in Figure~\ref{fig:price_raw_dropoff}.} }
    \label{fig:price_bias_dropoff}
\end{figure*}

Figure~\ref{fig:price_bias_pickup} and Figure~\ref{fig:price_bias_dropoff} display effect size charts of AI bias in fare pricing associated with trip frequency, ethnicity, age, education, poverty level, citizenship, and home value in Chicago neighborhoods. 
Figure~\ref{fig:price_bias_pickup} displays the effect sizes calculated for a neighborhood when it is the origin of a trip.
Figure~\ref{fig:price_bias_dropoff} displays the effect sizes calculated for a neighborhood when it is the destination of a trip.

Effect sizes of pickup and dropoff frequency on fare price suggest that neighborhoods with lower trip frequencies also have lower fares.
Trip frequencies represent ride requests that were accepted by ridehailing applications, or instances where consumer demand was met.
However riders whose trips were not accepted are not represented in this dataset, which is why trip frequency only accounts for a portion of demand.
Still, the positive relationship between the visible demand in this dataset (trip frequency) and fare price, shows that neighborhoods with higher demand have higher fare prices.
From Figures~\ref{fig:price_bias_pickup} and~\ref{fig:price_bias_dropoff}, charts ``A = Pickup Frequency / $m^2$'' and  ``A = Dropoff Frequency / $m^2$'' show that effect sizes decrease as pickup and dropoff frequencies in a census tract increase.
This means that the price is always lower for neighborhoods with lower pickup and dropoff frequencies when compared to neighborhoods with higher pickup and dropoff frequencies. 
This is consistent whether considering a neighborhood the origin of a trip, or the destination, shown by the similar trends for these two charts across both Figures~\ref{fig:price_bias_pickup} and~\ref{fig:price_bias_dropoff}.

Effect sizes of the non-white percentage in a neighborhood and fare price show that neighborhoods that on average, neighborhoods that are less white have higher fare prices than neighborhoods that are more white.
Considering effect size chart ``A = \% Non-white'' in Figures~\ref{fig:price_bias_pickup} and~\ref{fig:price_bias_dropoff}, effect sizes increase as the percentage of non-white population in a neighborhood increase to about $40\%$, and then effect sizes level out.
However, because all effect sizes are negative, Figures~\ref{fig:price_bias_pickup} and~\ref{fig:price_bias_dropoff} show that the difference between average fare price of a more non-white neighborhood is always higher than the average fare price of a less non-white neighborhood.
This is consistent regardless of whether neighborhoods are considered as the origin or the destination of a trip. 

Looking at effect size chart ``A = \% Older than 40'' in Figure~\ref{fig:price_bias_pickup}, almost all effect sizes are positive, meaning being picked up in neighborhoods with a younger population generally always costs more than being picked up in neighborhoods with an older population. 
The only case when neighborhoods with more older residents pay more are when the percentage of population older than $40$ is less than approximately 10\%.
However the combined effect size (see Table~\ref{table:price_table_ext}) is negative, reflecting the overall trend of cheaper prices in neighborhoods with more residents over $40$.
The same results can be seen when being dropped off in neighborhoods with younger populations versus neighborhoods with older populations, shown in Figure~\ref{fig:price_bias_dropoff}.

Fare prices decrease as the percentage of population in a neighborhood that has a high school diploma or less increases.
Fare prices increase as the percentage of population in a neighborhood below the poverty line increases.
Effect size charts ``\% High School Education or Less'' and ``\% Below Poverty Line'' in Figures~\ref{fig:price_bias_pickup} and~\ref{fig:price_bias_dropoff} have similar decreasing effect sizes regardless of whether examining these demographics at the origin or destination of a trip.
However the effect sizes for ``\% High School Education or Less'' are positive until $A \approx 0.70$. 
This shows that average fare price is generally more expensive for neighborhoods with high levels of college education, confirmed by Table~\ref{table:price_table_ext}. 
The effect size in the ``\% Below Poverty Line'' chart becomes negative approximately when $A \approx 0.15$.
This shows that average fare price is generally more expensive for neighborhoods with high levels of their population living below the poverty line, confirmed by Table~\ref{table:price_table_ext}.
These trends are consistent for both demographic attributes (high school education and poverty level) regardless of whether the attributes are being considered at trip origin or destination.

Fare price does not change significantly with the percentage of U.S. citizens in a neighborhood.
Fare prices decrease as the percentage of population living in homes below the median house price in a neighborhood increases.
Effect size charts ``\% Non-U.S. Citizens'' and ``\% Below Median House Price'' in Figures~\ref{fig:price_bias_pickup} and~\ref{fig:price_bias_dropoff} both show little variance regardless of the interval being examined. 
The effect sizes in ``\% Non-U.S. Citizens'' chart remains at approximately $\sim0.0$, which is reflected in the combined effect sizes for the demographic (see Table~\ref{table:price_table_ext}), which are all insignificant.
This means there is no significant association between the size of the immigrant population in a neighborhood and fare prices.
The effect sizes in chart ``\% Below Median House Price'' remains consistently positive, meaning that fare prices for neighborhoods with low home values are always lower than neigborhoods with high home values.
Effect sizes for both demographic attributes (non-U.S. citizen, below median house price) are consistent regardless of whether the attributes are being considered at trip origin or destination.

\begin{figure*}
    \centering
    \includegraphics[width=\textwidth]{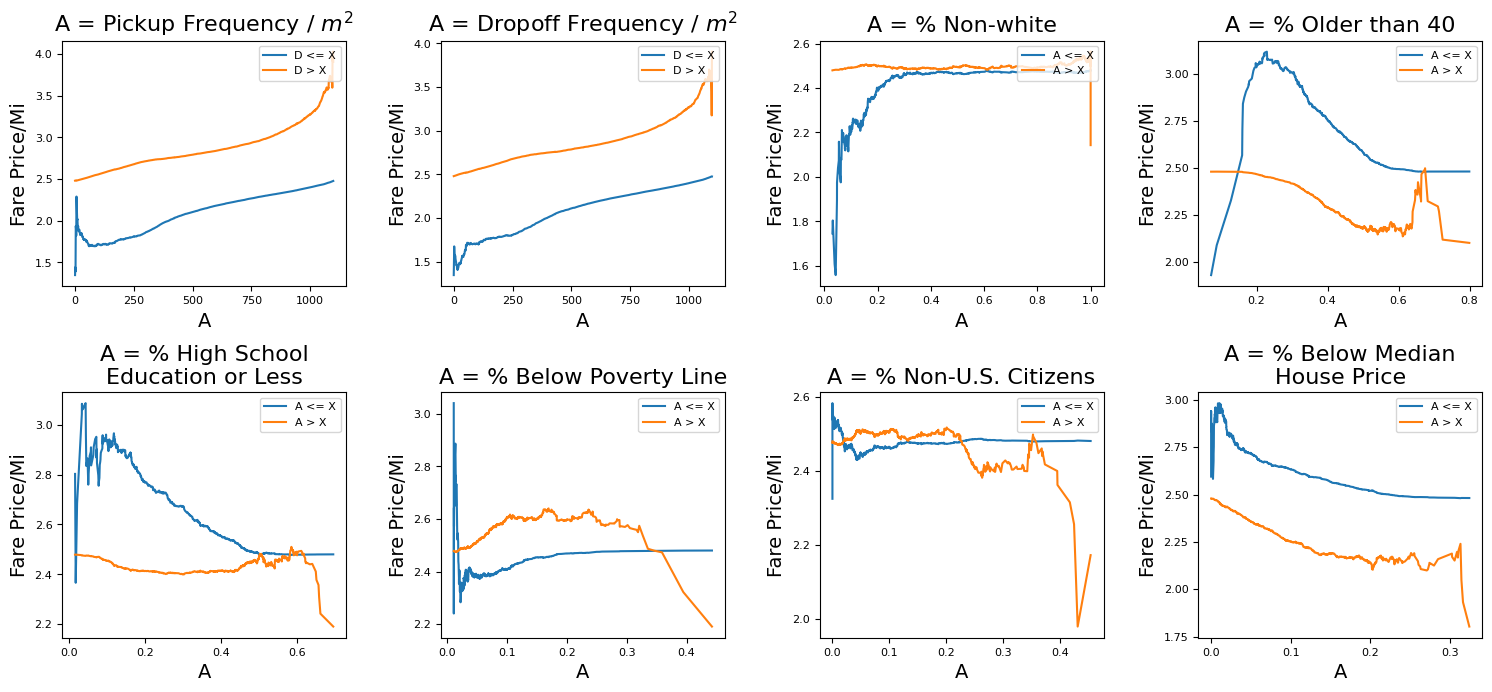}
    \caption{Raw price differences at pickup location - The change in fare price per mi for neighborhoods with $A <= X$ and $A > X$ is shown as the value of neighborhood attribute $A$ is increased. $X$ refers to the values on the x axis of each chart.
    Raw price differences are calculated for pickup and dropoff frequencies per neighborhood. 
    Raw price differences are also calculated for the percentage of non-white, over $40$, high school educated or less, below poverty line, and non-U.S. citizen populations, as well as the percentage of people living in homes below median home value in Chicago.
    }
    \label{fig:price_raw_pickup}
\end{figure*}

\begin{figure*}
    \includegraphics[width=\textwidth]{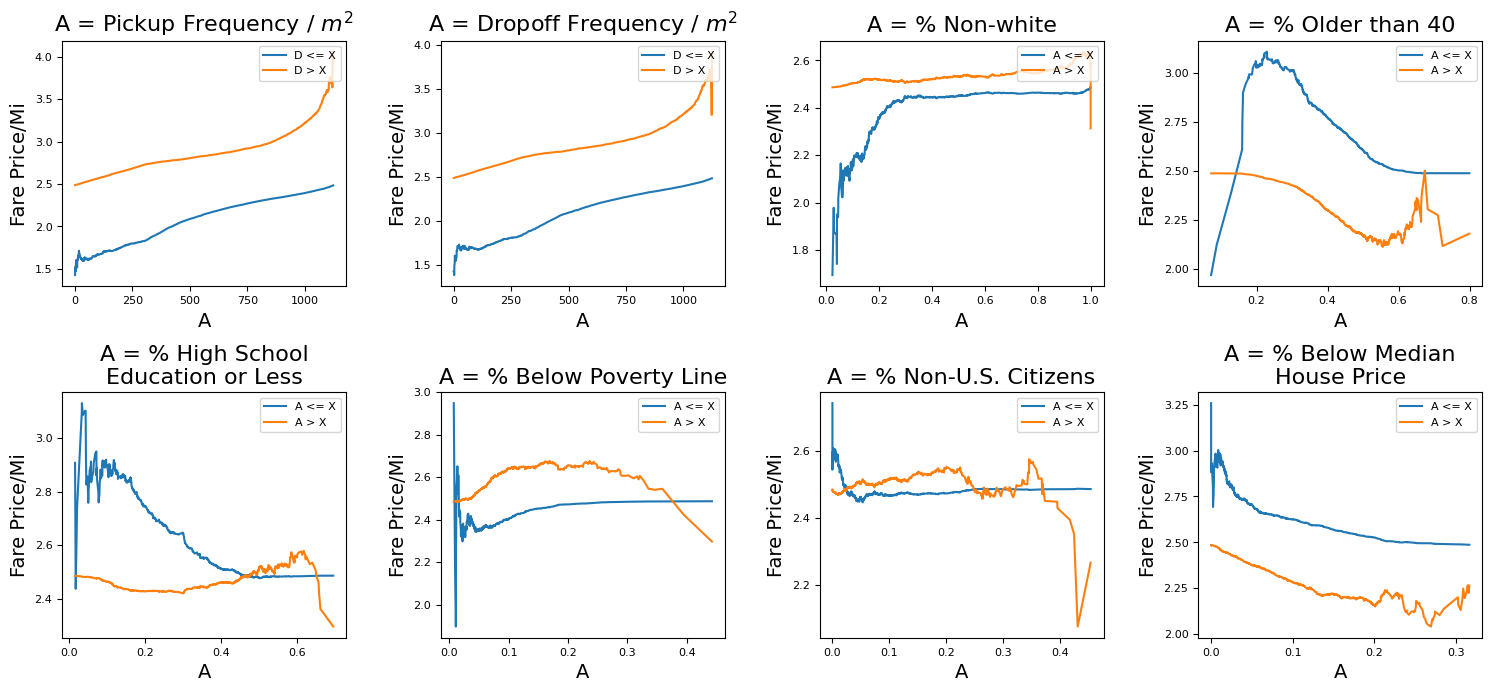}
    \caption{Raw price differences at dropoff location - The change in fare price per mi for neighborhoods with $A <= X$ and $A > X$ is shown as the value of neighborhood attribute $A$ is increased. $X$ refers to the values on the x axis of each chart.
    Raw price differences are calculated for pickup and dropoff frequencies per neighborhood. 
    Raw price differences are also calculated for the percentage of non-white, over $40$, high school educated or less, below poverty line, and non-U.S. citizen populations, as well as the percentage of people living in homes below median home value in Chicago.
    }
    \label{fig:price_raw_dropoff}
\end{figure*}

Figure~\ref{fig:price_raw_pickup} and Figure~\ref{fig:price_raw_dropoff} display the difference in average fare price per mile for two groups of neighborhoods at various demographic intervals. 
Figure~\ref{fig:price_bias_pickup} displays the differences in average fare price per mile for neighborhoods when they are the origin of a trip.
Figure~\ref{fig:price_bias_dropoff} displays the differences in average fare price per mile for neighborhoods when they are the destination of a trip.
These figures can be used to determine the actual fare price difference per mile at specific demographic intervals.
For example, in Figure~\ref{fig:price_bias_pickup}, looking at the ``A = \% Older than 40'' chart, when $A=0.20$ the average fare price per mile for neighborhoods with greater a than $20\%$ population older than $40$ is approximately $\$2.50$, represented in orange.
The average fare price per mile for neighborhoods with less than a $20\%$ population older than $40$ is approximately $\$3.00$, represented in blue.

\subsection*{Extended Results: AI Bias on Fare Pricing with ACS Error Margins}
ACS demographic statistics are estimated values, calculated using voluntary surveys taken throughout the United States~\cite{acs1}.
In addition to demographic estimates, ACS also provides standard error margins indicating how far estimates may deviate from ground truth demographic statistics.
In order to account for the margin of error in ACS demographic estimates, all combined effect size calculations were repeated using minimum possible demographic estimates as well as maximum possible demographic estimate.
Minimum possible demographic estimates were calculated by subtracting the error margin from each demographic statistic, and maximum possible demographic estimates were calculated by adding the error margin to each demographic statistic.
These combined effect sizes for each demographic are shown in Table~\ref{table:price_table_ext}.
Combined effect sizes calculated using minimum and maximum demographic estimates are shown in the ``Min Est'' and ``Max Est'' columns in Table~\ref{table:price_table_ext}.

\begin{table*}[t]
    \centering
    \begin{tabular}{l|c|c|c|c|c|c}
    \multicolumn{7}{c}{\textbf{Pickup}}                                                                                                                            \\\hline 
                                              & \multicolumn{2}{c|}{\textbf{Min Estimate}} & \multicolumn{2}{c|}{\textbf{Estimate}} & \multicolumn{2}{c}{\textbf{Max Estimate}} \\
    \textbf{Attribute}                        & \textbf{Combined}        & \textbf{p}              & \textbf{Combined} & \textbf{p}          & \textbf{Combined} & \textbf{p}              \\\hline
                            & \textbf{Effect Size (d)} &               & \textbf{Effect Size (d)} &           & \textbf{Effect Size (d)} &               \\\hline
    Pickup Frequency / $m^2$                  & -1.57      & $<10^{-3}$              & -1.57      & $<10^{-3}$          & -1.58      & $<10^{-3}$              \\      
    Dropoff Frequency / $m^2$                 & -1.57      & $<10^{-3}$              & -1.57      & $<10^{-3}$          & -1.57      & $<10^{-3}$              \\      
    \% Non-white                              & -0.13      & 0.20                    & -0.22      & 0.02                & -0.23      & 0.02                    \\      
    \% Older than 40                          &  0.64      & $<10^{-3}$              &  0.66      & $<10^{-3}$          &  0.67      & $<10^{-3}$              \\      
    \% High School Education or Less          &  0.34      & $<10^{-3}$              &  0.24      & 0.01                &  0.30      & $<10^{-3}$              \\      
    \% Below Poverty Line                     & -0.11      & 0.31                    & -0.19      & 0.05                & -0.08      & 0.45                    \\      
    \% Non-U.S. Citizens                      & -0.07      & 0.51                    & -0.10      & 0.35                &  0.01      & 0.93                    \\      
    \% Below Median House Price               &  0.24      & 0.01                    &  0.23      & 0.02                &  0.27      & 0.02                    \\\\
    \multicolumn{7}{c}{\textbf{Dropoff}}                                                                                                                            \\\hline 
                                              & \multicolumn{2}{c|}{\textbf{Min Estimate}} & \multicolumn{2}{c|}{\textbf{Estimate}} & \multicolumn{2}{c}{\textbf{Max Estimate}}  \\
    \textbf{Attribute}                        & \textbf{Combined}        & \textbf{p}              & \textbf{Combined} & \textbf{p}          & \textbf{Combined} & \textbf{p}              \\\hline
                            & \textbf{Effect Size (d)} &               & \textbf{Effect Size (d)} &           & \textbf{Effect Size (d)} &               \\\hline
    Pickup Frequency / $m^2$                  & -1.59      & $<10^{-3}$              & -1.59      & $<10^{-3}$          & -1.59      & $<10^{-3}$              \\      
    Dropoff Frequency / $m^2$                 & -1.57      & $<10^{-3}$              & -1.57      & $<10^{-3}$          & -1.57      & $<10^{-3}$              \\      
    \% Non-white                              & -0.29      & $<10^{-3}$              & -0.32      & $<10^{-3}$          & -0.35      & $<10^{-3}$              \\      
    \% Older than 40                          &  0.66      & $<10^{-3}$              &  0.69      & $<10^{-3}$          &  0.70      & $<10^{-3}$              \\      
    \% High School Education or Less          &  0.25      & $<10^{-3}$              &  0.15      & 0.17                &  0.19      & 0.06                    \\      
    \% Below Poverty Line                     & -0.18      & 0.07                    & -0.28      & $<10^{-3}$          & -0.21      & 0.04                    \\      
    \% Non-U.S. Citizens                      & -0.01      & 0.93                    & -0.07      & 0.54                &  0.03      & 0.79                    \\      
    \% Below Median House Price               &  0.28      & $<10^{-3}$              &  0.19      & 0.06                &  0.19      & 0.06                    \\      
\end{tabular}
\caption{Combined effect sizes on fare price per mile by neighborhood attributes - combined effect sizes (column ``Effect size'') are shown for the fare price per mile given a set of neighborhood attributes. 
``Pickup'' and ``Dropoff'' columns designate fare price per mile when being picked up or dropped of in a neighborhood and ``p'' presents the p-value for effect size calculations. 
ACS data contains estimates for demographic statistics by census tract. 
Additionally, ACS provides a standard error margin for the demographic statistics estimates. 
Results calculated using the estimate with the error subtracted, as well as the estimate with the error added are shown in the ``Min Estimate'' and ``Max Estimate'' columns. 
}
\label{table:price_table_ext}
\end{table*}

\end{document}